\documentclass[twocolumn,english,aps,prb,superscriptaddress]{revtex4}
\usepackage{amsmath}
\usepackage{graphicx}
\usepackage{amssymb}

\makeatletter

\@ifundefined{textcolor}{}
{%
 \definecolor{BLACK}{gray}{0}
 \definecolor{WHITE}{gray}{1}
 \definecolor{RED}{rgb}{1,0,0}
 \definecolor{GREEN}{rgb}{0,1,0}
 \definecolor{BLUE}{rgb}{0,0,1}
 \definecolor{CYAN}{cmyk}{1,0,0,0}
 \definecolor{MAGENTA}{cmyk}{0,1,0,0}
 \definecolor{YELLOW}{cmyk}{0,0,1,0}
 }

\usepackage{babel}

\makeatother

\usepackage{babel}

\begin{document}

\title{Interface energy of two band superconductors }

\author{Jani Geyer}

\affiliation{Department of Physics and Astronomy and Ames Laboratory, Iowa State
University, Ames, Iowa 50011, USA }

\affiliation{Department of Physics, University of Stellenbosch, Stellenbosch 7600,
South Africa}

\affiliation{National Institute for Theoretical Physics (NITheP), Stellenbosch,
Private Bag X1, Matieland, 7602, South Africa}

\author{Rafael M. Fernandes}

\affiliation{Department of Physics and Astronomy and Ames Laboratory, Iowa State
University, Ames, Iowa 50011, USA }

\author{V. G. Kogan}

\affiliation{Department of Physics and Astronomy and Ames Laboratory, Iowa State
University, Ames, Iowa 50011, USA }

\author{J\"{o}rg Schmalian}

\affiliation{Department of Physics and Astronomy and Ames Laboratory, Iowa State
University, Ames, Iowa 50011, USA }

\date{\today}
\begin{abstract}
Using the Ginzburg-Landau theory for two-band superconductors, we
determine the surface energy $\sigma_{s}$ between coexisting normal
and superconducting\textbf{\ }states at the thermodynamic critical
magnetic field. Close to the transition temperature, where the Ginzburg-Landau
theory is applicable, we demonstrate that the
two-band problem maps onto an effective single band problem. While
the order parameters of the two bands may have different amplitudes
in the homogeneous bulk, near $T_{c}$ the Josephson-like coupling
between the bands leads to the same spatial dependence of both order
parameters near the interface. This finding puts into question the
possibility of intermediate, so called \emph{type-1.5} superconductivity,
in the regime where the Ginzburg-Landau theory applies. 
\end{abstract}
\maketitle

\section{Introduction}

Depending on the behavior in external magnetic fields, superconductors
are classified as type-I or type-II. \ In type I superconductors,
the surface energy density $\sigma_{s}$ between regions of finite
and zero order parameters, coexisting at the thermodynamic critical
field $H_{c}$, is positive. \cite{Ginzburg50} In type-II superconductors
\
this energy is negative and a homogeneous superconducting state is
no longer stable, leading to the formation of a vortex lattice.\citep{Abrikosov57}
Within the Ginzburg-Landau (GL) theory \cite{Ginzburg50} for one-band
superconductors the interface energy per unit area, \begin{equation}
\sigma_{s}=\lambda\frac{H_{c}^{2}}{4\pi}\Upsilon\left(\kappa\right),\label{sigmas}\end{equation}
 is determined by the value of the thermodynamic critical field, $H_{c}$,
the magnetic penetration depth, $\lambda$, and the dimensionless
function, $\Upsilon\left(\kappa\right)$, that depends on the GL parameter
$\kappa=\lambda/\xi$, the ratio of the penetration depth and the
superconducting coherence length. Properties of $\Upsilon\left(\kappa\right)$
are discussed, e.g. in Ref. \onlinecite{SaintJ69}. In the regimes
of extreme type-I and type-II superconductivity \begin{equation}
\Upsilon\left(\kappa\right)=\left\{ \begin{array}{cc}
\frac{2^{3/2}}{3}\kappa^{-1} & \text{ if \ }\kappa\ll1\\
-\frac{4}{3}\left(\sqrt{2}-1\right) & \text{ if \ }\kappa\gg1\end{array}\right..\label{uni func}\end{equation}

\begin{figure}
 \includegraphics[width=0.9\columnwidth]{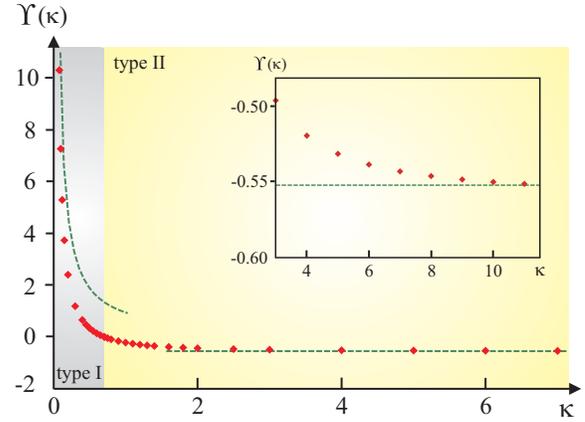}
\caption{(Color online) The function, $\Upsilon\left(\kappa\right)$, of the
one-band problem calculated numerically (full points) compared to
theoretical limit of Eq. \protect\ref{uni func} (dashed lines).
The inset shows an enlargement of the large $\kappa$ domain. }

\end{figure}
We have evaluated this function numerically and the result is shown
in Fig. 1. The transition between type-I and type-II behavior occurs
for $\kappa=2^{-1/2}$, where $\Upsilon\left(\kappa\right)$ changes
sign. 

Fermi surfaces in many superconductors may consist of two or more
well separated sheets with different energy gaps.\cite{Suhl59,Moskalenko59}
Evidence for two energy gaps was obtained in high-purity superconducting
Nb, Ta, and V,\citep{Shen} and Nb-doped SrTiO$_{3}$.\citep{Binnig}
Recently, tunneling\citep{Giubileo01,Iavarone02, Eskildsen02} and
point contact\citep{Szabo01,Schmidt01} spectroscopies, as well as
heat capacity measurements\citep{Wang01,Bouquet01,Yang01} for MgB$_{2}$\cite{MgB2E01,MgB2E02,MgB2T01,MgB2T02,Askerzade02}
give clear evidence for two-band superconductivity with gaps $\Delta_{1}\simeq0.7\mathrm{meV}$
and $\Delta_{2}\simeq2.5\mathrm{meV}$ (for recent reviews, see Refs.
\onlinecite{Canfield03} and \onlinecite{Xi08}). Other systems that
have been discussed as two-band systems are RNi$_{2}$B$_{2}$C with
R=Lu,Y,\citep{Heinecke95,Schulga95} 2H-NbSe$_{2}$\citep{Yokoya01}
and the recently discovered FeAs superconductors.\citep{Ding09} In
all cases the amplitude of the superconducting gap is different for
different sheets of the Fermi surface. 

Motivated by the study of these multi-band superconductors, the term
\emph{type-1.5} superconductivity has been coined\citep{Moshchalkov09}
to emphasize the possibility of a state that is intermediate between
the two regimes. Specifically, one considers two-component or two-band
systems with order parameters $\Psi_{1}\left(\mathbf{r}\right)$ and
$\Psi_{2}\left(\mathbf{r}\right)$ that have qualitatively different
spatial dependence, with different respective coherence lengths $\xi_{1}$
and $\xi_{2}$. The existence of these two length scales emerges from
the assumption that one can neglect the Josephson type coupling between
two order parameters.\ The regime where one expects novel behavior
is obviously the limit $\xi_{1}\ll\lambda\ll\xi_{2}$. Then one order
parameter component might behave as a type-I superconductor while
the other follows the type-II behavior. Consequences of such behavior
were discussed in Ref. \onlinecite{Babaev05} where it was concluded
that properties emerge that fall outside the usual type-I/type-II
dichotomy. For example, the emergence of ``vortex molecules''
and of an inhomogeneous state comprising a mixture of domains of a
two-component Meissner state and vortex clusters were proposed. \ In
Ref. \onlinecite{Wang09} the surface energy for such a system was
analyzed with the conclusion that $\Upsilon\left(\kappa\right)$ must
be replaced by a function that depends on several dimensionless quantities,
in particular on the ratio $\xi_{1}/\xi_{2}$. Changing $\xi_{1}/\xi_{2}$
at fixed penetration depth was shown to yield a sign change of $\sigma_{s}$.
\ 

Obviously, even in multiband superconductors, the sign of the surface
energy is either positive or negative. Thus, it seems more appropriate
to discuss the physics that was investigated in Ref. \onlinecite{Moshchalkov09}
within the GL approach,\cite{Askerzade06} as interesting modifications
of type-II superconductivity. More importantly, it is crucial to analyze
what exactly happens in a multiband superconductor in the vicinity
of the transition temperature, with \begin{equation}
\tau=\left(T_{c}-T\right)/T_{c}\ll1,\end{equation}
in the regime where the GL approach is valid (ignoring, as usual,
critical fluctuations).

One of the key features of the two-band GL model is the Josephson like coupling between the two bands, 
\begin{equation}
f_{c}\left(\mathbf{r}\right)=-\eta\left(\Psi_{1}^{\ast}\left(\mathbf{r}\right)\Psi_{2}\left(\mathbf{r}\right)+\Psi_{2}^{\ast}\left(\mathbf{r}\right)\Psi_{1}\left(\mathbf{r}\right)\right),\label{Fc}\end{equation}
 in the expansion of the GL free energy density. Refs. \onlinecite{Babaev05}
and \onlinecite{Wang09} analyze the limit $\eta=0$, but assume that
both order parameters, while uncoupled, order at the same temperature.
The more realistic regime is clearly the one where the common transition
temperature is the consequence of a finite order parameter coupling
$\eta$.

In this paper we determine the surface energy $\sigma_{s}$ of a two-band
GL model including the coupling, Eq.(\ref{Fc}), between the bands,
i.e. we consider $\eta\neq0$. We find that in the regime $\tau\ll1$,
where the GL theory provides the correct mean field description, Eq.(\ref{sigmas})
continues to be the correct expression for the interface energy with
same function $\Upsilon\left(\kappa\right)$,\textbf{\ }which implies
that the surface energy continues to change sign\textbf{\ }at $\kappa=2^{-1/2}$.
The multi-component nature of the order parameter enters the GL $\kappa$
through the values of $\lambda$ and\begin{equation}
\xi=\left(\xi_{1}^{-2}+\xi_{2}^{-2}\right)^{-1/2}.\label{xi}\end{equation}
 A detailed definition of $\lambda$ and $\xi_{i}\,$in the two-band
problem is presented\textbf{\ }below. We also find that\textbf{,}
while the order parameters may have different amplitudes in the homogeneous
bulk, $\Psi_{1,0}$ and $\Psi_{2,0}$, close to the transition temperature,
i.e. for small $\tau$, they have the same spatial dependence near
the interface. In particular: \begin{equation}
\frac{\Psi_{1}\left(z\right)}{\Psi_{2}\left(z\right)}\mathbf{=}\frac{\Psi_{1,0}}{\Psi_{2,0}}+\mathcal{O}\left(\tau\right),\label{ops}\end{equation}
i.e. the coupling enforces the same spatial dependence for both components.
\ $\Psi_{1}\left(\mathbf{r}\right)$ and $\Psi_{2}\left(\mathbf{r}\right)$
vary in space on the single length scale $\xi$ of Eq.(\ref{xi}).
Close to a superconducting transition it is then sufficient to introduce
only one order parameter to characterize the symmetry broken state.
An exception is the case where two completely uncoupled order parameters
are accidentally degenerate, i.e. both components accidentally have
the exact same transition temperatures $T_{c}$, while they have,
at the same time, different coherence lengths. This is the scenario
considered in Refs. \onlinecite{Babaev05} and \onlinecite{Wang09}.
We stress, that our results do not preclude novel type-II behavior
that may occur deeper in the ordered state away of the GL domain.
This is however beyond the limit of applicability of the GL theory.
In the next section we present our analysis. We summarize our findings
in section 3.

\section{Two-band Ginzburg-Landau Theory}

We start with the free energy, \begin{equation}
F=\int f\left(\mathbf{r}\right)d^{3}r,\label{FGL}\end{equation}
 of a two-band superconductor. $F$ is a functional of the pairing
wave functions $\Psi_{1}$ and $\Psi_{2}$ of the two components or
bands and of the vector potential $\mathbf{A}$ associated with the
magnetic field $\mathbf{B=\nabla\times A}$. The free energy density,
$f\left(\mathbf{r}\right)$, relative to the zero field normal state
value, is: \begin{equation}
f\left(\mathbf{r}\right)=f_{1}\left(\mathbf{r}\right)+f_{2}\left(\mathbf{r}\right)+f_{c}\left(\mathbf{r}\right)+\frac{B^{2}\left(\mathbf{r}\right)}{8\pi}.\label{FGL2}\end{equation}
 Here the $f_{j}\left(\mathbf{r}\right)$, with $j=1,2$, are the
GL expansions of the two bands: \begin{equation}
f_{j}=a_{j}\left\vert \Psi_{j}\right\vert ^{2}+\frac{1}{2}b_{j}\left\vert \Psi_{j}\right\vert ^{4}+\frac{1}{2m_{j}^{\ast}}\left\vert \left(\frac{\hbar}{i}\nabla-\frac{e^{\ast}}{c}\mathbf{A}\right)\Psi_{j}\right\vert ^{2},\label{FGL3}\end{equation}
 and $\ f_{c}\left(\mathbf{r}\right)$ is the coupling term given
in Eq.(\ref{Fc}). Here\textbf{ $b_{j}>0$} and the bands' effective
masses are $m_{j}$. The physical values of the order parameter and
vector potential are determined via $\delta F/\delta\Psi_{i}=\delta F/\delta A_{\alpha}=0$.
In principle additional coupling terms such as $\left(\Psi_{1}^{\ast}\Psi_{2}\right)^{2}$
, $\nabla\Psi_{1}^{\ast}\cdot\nabla\Psi_{2}$, etc. are allowed. For
clean multiband systems, a weak coupling expansion yields that \ the
coefficients of such terms vanish due to momentum conservation\citep{Zhitomirsky04}.
In addition, even if present, such terms are sub-leading close to
the transition temperature point when compared to $f_{c}\left(\mathbf{r}\right)$
of Eq.(\ref{Fc}).

We first discuss the homogeneous, zero field solution. Ignoring the
inter-band coupling, $f_{c}$, one finds, as usual, $\Psi_{i,0}\left(\eta=0\right)=\sqrt{-a_{i}/b_{i}}$
for $a_{i}<0$ and $\Psi_{i,0}\left(\eta=0\right)=0$ for $a_{i}>0$.
In the general case of $f_{c}\neq0$, however, the common critical
temperature $T_{c}$ is not equal to either of $T_{c,j}$ and there
is no reason that both coefficients $a_{i}\left(T\right)$ change
sign at the same temperature. At $T_{c}$ the smallest eigenvalue
of the matrix of the homogeneous quadratic terms in $f\left(\mathbf{r}\right)$
vanishes. In our problem, this eigenvalue is \begin{equation}
r_{-}=\frac{1}{2}\left(a_{1}+a_{2}-\sqrt{\left(a_{1}-a_{2}\right)^{2}+4\eta^{2}}\right);\end{equation}
it vanishes for $\eta^{2}=a_{1}\left(T_{c}\right)a_{2}\left(T_{c}\right)$.
Thus, it must hold that $a_{1,2}\left(T_{c}\right)>0$, as $r_{-}$
would be negative if one of the two $a_{i}$ is smaller or equal to
zero. Thus, close to $T_{c}$, $a_{i}>0$ and the interband coupling
enhances the transition temperature compared to the largest of the
$T_{c,j}$ for the $\eta=0$ limit. 

To proceed, we introduce the dimensionless ratio \begin{equation}
t\equiv\frac{\eta^{2}-a_{1}a_{2}}{a_{1}a_{2}}\propto\frac{T_{c}-T}{T_{c}},\label{t}\end{equation}
 that vanishes at $T_{c}$ (see also the Appendix). It is convenient
to eliminate $\eta^{2}=\left(1+t\right)a_{1}a_{2}$ in favor of $t$.
Thus, small $t$ naturally corresponds to finite interband coupling
$\eta$. For small $t$ we have $r_{-}\simeq-ta_{1}a_{2}/\left(a_{1}+a_{2}\right)$
and the smallest eigenvalue changes sign at $t=0$.

The free energy minimization of the homogeneous problem for $\eta\neq0$
leads to the system \begin{equation}
\begin{array}{c}
a_{1}\Psi_{1}+b_{1}\Psi_{1}^{3}-\eta\Psi_{2}=0\\
a_{2}\Psi_{2}+b_{2}\Psi_{2}^{3}-\eta\Psi_{1}=0\end{array},\end{equation}
which is readily reduced to a fourth order equation for $\Psi_{1}^{2}$
that can be solved using known formulas for the roots of a quartic
equation. One can simplify the problem by recognizing that the GL
theory is only valid in the vicinity of the transition temperature,
$t\ll1$. The homogeneous order parameters can easily be determined
to leading order in $t$: \begin{eqnarray}
\Psi_{1,0}^{2} & = & u_{1}t\ \text{\ with }u_{1}=\frac{a_{2}^{2}a_{1}}{a_{2}^{2}b_{1}+a_{1}^{2}b_{2}},\label{OPstat1}\\
\Psi_{2,0}^{2} & = & u_{2}t\ \text{\ with }u_{2}=\frac{a_{1}^{2}a_{2}}{a_{2}^{2}b_{1}+a_{1}^{2}b_{2}},\label{OPstat}\end{eqnarray}
 where the subscript $0$ is to denote the zero-field solution. Hence,
the temperature dependence of the order parameters is as expected:
\begin{equation}
\Psi_{j,0}^{2}\propto t\propto\frac{T_{c}-T}{T_{c}}.\end{equation}
 We stress that within GL theory there is no reason to go to terms
of higher orders in $t$. Of course, away from $T_{c}$ corrections
can be significant, in particular for small $\eta$, but those effects
require a microscopic approach based on Bogoliubov-de Gennes equations.\citep{DeGennes}

Close to\textbf{\ $T_{c}$}, we can also determine the thermodynamic
critical field by imposing\textbf{\ }$f\left(H_{c}\right)=0$: \begin{equation}
\frac{H_{c}^{2}}{4\pi}=\sum_{j=1}^{2}b_{j}\left\vert \psi_{j}\right\vert ^{4}=\frac{a_{1}^{2}a_{2}^{2}t^{2}}{a_{2}^{2}b_{1}+a_{1}^{2}b_{2}}.\end{equation}
 One can \textit{formally} define the one-band penetration depth as
$\lambda_{i}^{-2}=4\pi e^{\ast2}\Psi_{i,0}^{2}/\left(m_{i}^{\ast}c^{2}\right)$.
Since the additive superfluid density is proportional to $\lambda^{-2}$,
the actual London penetration depth is \begin{equation}
\lambda^{-2}=\lambda_{1}^{-2}+\lambda_{2}^{-2}.\ \end{equation}
Using Eq.(\ref{OPstat}), we obtain \begin{equation}
\lambda^{-2}=\frac{4\pi e^{\ast2}a_{1}a_{2}}{c^{2}}\frac{a_{2}/m_{1}^{\ast}+a_{1}/m_{2}^{\ast}}{a_{2}^{2}b_{1}+a_{1}^{2}b_{2}}t.\end{equation}
 It is now straightforward to set up the formalism to determine the
interface energy.

\section{The interface energy}

In evaluating the surface energy we follow closely the classical approach
that was used for the single band problem. \cite{Ginzburg50} Consider
the interface between superconducting and normal half-spaces at the
plane $z=0$. The field $H$ is applied along the $x$ axis parrallel
to the interface and equal to $H_{c}$ to ensure coexistence of two
phases. Then the magnetic induction has only one component $B_{x}=B\left(z\right)$
and the vector potential can be chosen as $A_{y}=-A\left(z\right)$,
as shown in Fig. 2, yielding \begin{equation}
B\left(z\right)=A'\left(z\right).\end{equation}

\begin{figure}
 \includegraphics[width=0.8\columnwidth]{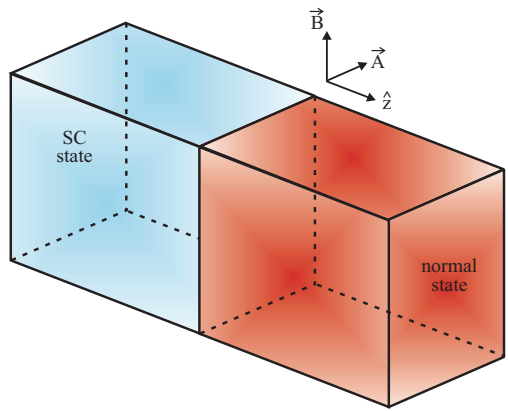}
\caption{(Color online) Schematic representation of the interface between the
normal and the superconducting state.}

\end{figure}

We can choose real order parameters that vary along the $z$-direction.
The Gibbs free energy per unit area that is minimum in a given applied
field reads:\begin{equation}
G=F-\frac{H_{c}}{4\pi}\int_{-\infty}^{\infty}dz\: B\left(z\right).\end{equation}
 Far away from the interface we have on one side the normal state
with $\Psi_{i}\left(z\rightarrow-\infty\right)\rightarrow0$ and $B\left(z\rightarrow-\infty\right)\rightarrow H_{c}$,
while on the other side for $z\rightarrow\infty$ we have the homogeneous
superconducting state with $B\rightarrow0$ and $\Psi_{i}\rightarrow\Psi_{i,0}$.

It is convenient to introduce dimensionless quantities: \begin{equation}
\psi_{j}^{2}=\frac{\Psi_{j}^{2}}{Y_{j}\tau}\text{, \ }b=\frac{B}{\sqrt{2}H_{c}}\text{, \ and }s=\frac{z}{\lambda},\end{equation}
 which imply that the dimensionless vector potential $a=A/\left(\sqrt{2}H_{c}\lambda\right)$.
The surface energy is then given by \begin{equation}
\sigma_{s}=\lambda\frac{H_{c}^{2}}{4\pi}\Sigma\left[\psi_{1},\,\psi_{2},\, a\right],\end{equation}
 where $\Sigma$ is a functional that must be minimized with respect
to the $\psi_{1}$, $\psi_{2}$ and $a$ to yield $\sigma_{s}$. After
simple algebra we obtain: \begin{equation}
\Sigma=\int\left(V\left(\psi_{1},\psi_{2},a\right)+\sum_{i}\kappa_{i}^{-2}\psi_{i}^{\prime2}+\left(a^{\prime}-2^{-1/2}\right)^{2}\right)ds\label{Sigma functional}\end{equation}
 where \begin{eqnarray}
V\left(\psi_{1},\psi_{2},a\right) & = & \frac{\psi_{1}^{2}+\psi_{2}^{2}-2\sqrt{1+t}\psi_{1}\psi_{2}}{t}\ +\frac{u}{2}\psi_{1}^{4}\notag\label{V}\\
 &  & +\frac{1-u}{2}\psi_{2}^{4}+\frac{\kappa_{2}^{2}\psi_{1}^{2}+\kappa_{1}^{2}\psi_{2}^{2}}{\kappa_{1}^{2}+\kappa_{2}^{2}}a^{2}.\end{eqnarray}
We use here the following notations: $\psi_{i}^{\prime}=d\psi_{i}/ds$,
$b=a^{\prime}=da/ds$, $u=u_{1}$, as given in Eq.(\ref{OPstat1}),
and \begin{equation}
\kappa_{i}=\frac{\lambda}{\xi_{i}},\quad\mathrm{with}\quad\xi_{i}^{2}=\frac{\hbar^{2}}{2a_{i}m_{i}^{\ast}t}.\end{equation}
Minimization of $\Sigma$ gives a system of coupled differential equations
for $\psi_{i},\: a$: \begin{equation}
\frac{1}{\kappa_{i}^{2}}\psi_{i}^{\prime\prime}=\frac{1}{2}\frac{\partial V}{\partial\psi_{i}},\label{eqofmot1}\end{equation}

\begin{equation}
a^{\prime\prime}=\frac{1}{2}\frac{\partial V}{\partial a}.\label{eqofmot2}\end{equation}
Multiplying Eq.(\ref{eqofmot1}) by $\psi_{i}^{\prime}$, Eq.(\ref{eqofmot2})
by $a^{\prime}$ and summing, the first integral of this system
is obtained: \begin{equation}
\sum_{i}\kappa_{i}^{-2}\psi_{i}^{\prime2}+a^{\prime2}-V\left(\psi_{1},\psi_{2},a\right)=const.\end{equation}

The peculiar term in our analysis is the first one in $V\left(\psi_{1},\psi_{2},a\right)$
of Eq.(\ref{V}) that seems to be singular as $t\rightarrow0$. Expanding
for small $t$, we have:\begin{equation}
\frac{\psi_{1}^{2}+\psi_{2}^{2}-2\sqrt{1+t}\psi_{1}\psi_{2}}{t}\simeq\frac{\left(\psi_{1}-\psi_{2}\right)^{2}}{t}-\psi_{1}\psi_{2}.\end{equation}
 Thus, close to the transition temperature we must have $\psi_{1}=\psi_{2}$.
Introducing $\psi\left(s\right)=\psi_{1}\left(s\right)=\psi_{2}\left(s\right)$,
which is equivalent to Eq.(\ref{ops}), one obtains the surface energy
functional in the form: \begin{equation}
\Sigma=\int ds\left(V_{0}\left(\psi,a\right)+\kappa^{-2}\psi^{\prime2}+\left(a^{\prime}-2^{-1/2}\right)^{2}\right),\end{equation}
 with \begin{equation}
V_{0}\left(\psi,a\right)=-\psi^{2}\ +\frac{1}{2}\psi^{4}+\psi^{2}a^{2}\end{equation}
 and effective parameter $\kappa$ given by\begin{equation}
\kappa^{-2}=\kappa_{1}^{-2}+\kappa_{2}^{-2}.\label{kappatpt}\end{equation}
 This is an exact form of the functional for the standard one-band
surface enrgy problem.\cite{Ginzburg50,SaintJ69} 

It is worth noting that $\kappa_{i}$ enter the surface energy only
through the combination $\kappa$ of Eq.(\ref{kappatpt}). In particular,
this leads to Eq.(\ref{xi}) for the correlation length of the two
band problem with $\kappa=\lambda/\xi$. Thus, the interface problem
is identical to the one of a single band system, leading to Eq.(\ref{sigmas})
with the same function $\Upsilon\left(\kappa\right)$.

These conclusions are supported by numerical minimization of $\Sigma\left[\psi_{1},\psi_{2},a\right]$.
We discretized the interval $s=\left[0,2L\right]$ to $N$ equidistant
steps ($s_{j}=2jL/N$) and minimized $\Sigma$ with respect to $\psi_{1}\left(s_{i}\right)$,
$\psi_{2}\left(s_{i}\right)$ and $a\left(s_{i}\right)$ subject to
boundary conditions $a\left(0\right)=2^{-1/2}$, $a\left(2L\right)=0$,
and $\psi_{i}\left(0\right)=0$ and $\psi_{i}\left(2L\right)=\psi_{i,0}$.
The homogeneous bulk solutions $\psi_{i,0}$ approach the value $\psi_{i,0}=1$
for $t\rightarrow0$. Finally, since in the limit $2L\rightarrow\infty$
the interface position is arbitrary, at $z=L$ we assumed $\psi_{1}\left(L\right)=\frac{1}{2}$,
which centers the interface position in the large $\kappa$ limit. 

Our results for $N=400$ are shown in Figs. 3-6. In comparing Fig.
3 and Fig. 4, as well as Fig. 5 and Fig. 6, we show that the order
parameters do indeed approach the behavior with identical spatial
variation, as given by Eq.(\ref{ops}), as the critical temperature
is approached. 

In Figs. 3 and 4 we focus on the most nontrivial limit with $\kappa_{1}=0.45<2^{-1/2}<\kappa_{2}=5$.
Naively, one could expect $\psi_{1}$ to change on distances of the
order $\xi_{1}>\sqrt{2}\lambda$ (type-I behavior), while $\xi_{2}<\sqrt{2}\lambda$
suggests type-II behavior of $\psi_{2}$. Contrary to this expectation
we find that both order parameters are strongly coupled by the Josephson
energy and have increasingly similar spatial variation as $t\rightarrow0$.
As we will see below, the interface energy for this set of parameters
is positive and the system behaves as a type-I superconductor as $\kappa$
in Eq.(\ref{kappatpt}) is dominated by the smallest of the two $\kappa_{i}$. 

In Figs. 5 and 6 we show the behavior for $\kappa_{1}=3$ and $\kappa_{2}=4$,
corresponding indeed to a type-II superconductor with negative interface
energy (see below for explicit values). Again, both order parameters
follow the same spatial dependence and behave according to Eq.(\ref{ops})
as $t$ decreases. 

In addition, as $t$ decreases, the value of the minimized functional
of Eq.(\ref{Sigma functional}) approaches the value of the function
$\Upsilon\left(\kappa\right)$ of the single band problem with $\kappa$
determined by Eq.(\ref{kappatpt}). This can explicitly be seen in
the numerical results of $\Sigma_{min}$ and $\Upsilon\left(\kappa\right)$
corresponding to Figs. 3-6. The effective one-band solution with $\kappa=0.448$,
thus corresponding to Fig. 3 and Fig. 4, gives $\Upsilon\left(\kappa\right)=0.479$,
which differs from the numerical result of Fig. 3, $\Sigma_{min}=0.530$,
by $\sim11\%$. This difference decreases to $\sim2\%$ for a smaller
value of $t$ as shown in Fig. 4, for which $\Sigma_{min}=0.488$.
The numerical solutions shown in Fig. 5 obtained for $t=0.2$ correspond
to $\Sigma_{min}=-0.275$ whereas the effective one-band $\kappa=2.4$
yields $\Upsilon\left(\kappa\right)=-0.47$: hence $\Sigma_{min}$
and $\Upsilon$ differ by $\sim42\%$. Again, by decreasing the value
of $t$ to 0.01, we find results shown in Fig. 6 corresponding to
$\Sigma_{min}=-0.459$, now only by $\sim2\%$ different from $\Upsilon$.

\begin{figure}
 \includegraphics[width=1\columnwidth]{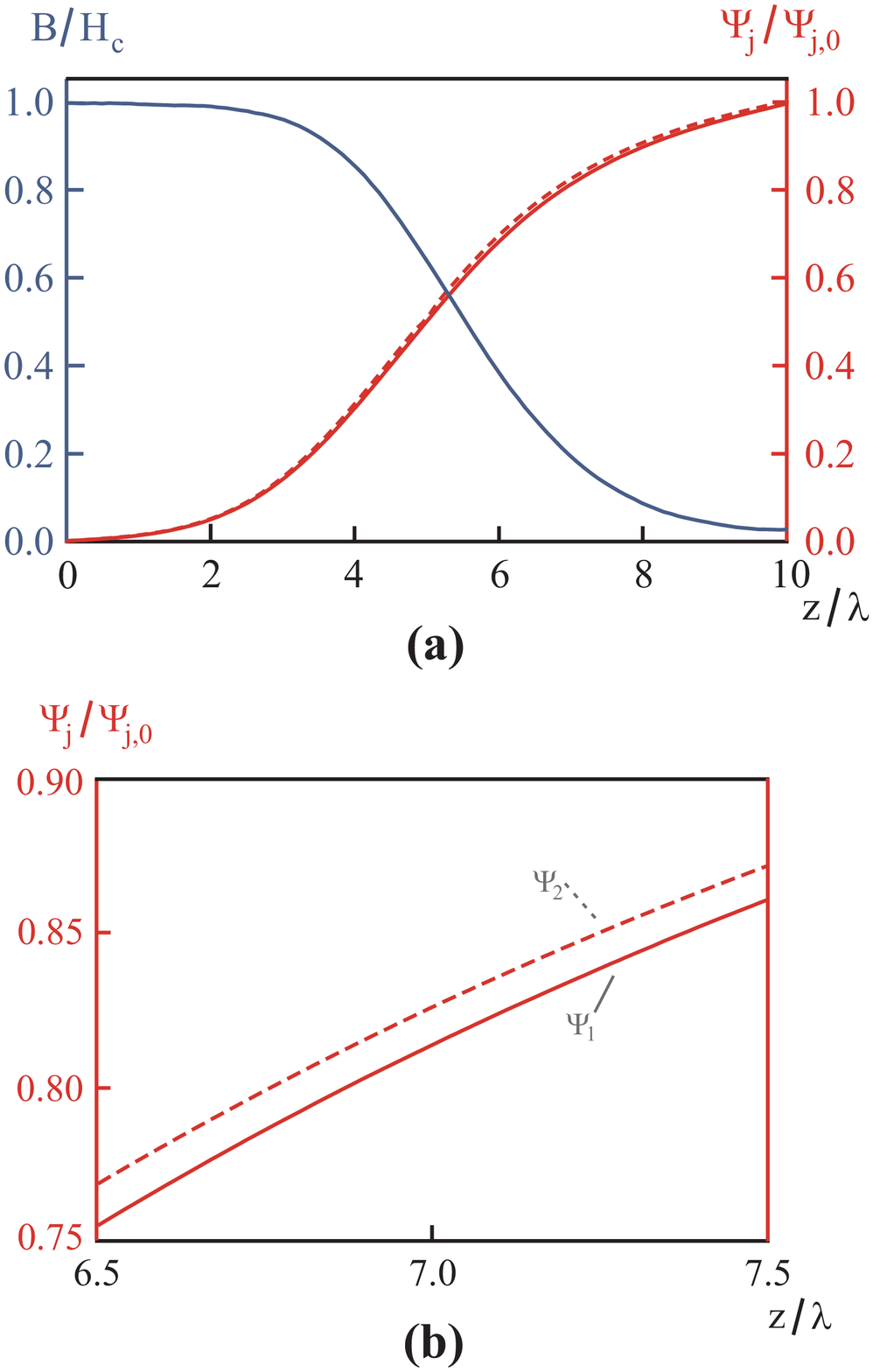}
\caption{(Color online) (a) The reduced field and order parameters obtained
numerically by minimizing the interface energy functional, $\Sigma$,
for $\kappa_{1}=0.45,\;\kappa_{2}=5,\; u=0.6,\; t=0.07$. The reduced
order parameter $\psi_{1}$ is shown by a solid line, the dashed line
is $\psi_{2}$. (b) The close-up of the order parameters for $6.5<z/\lambda<7.5$.}

\end{figure}

\begin{figure}
 \includegraphics[width=1\columnwidth]{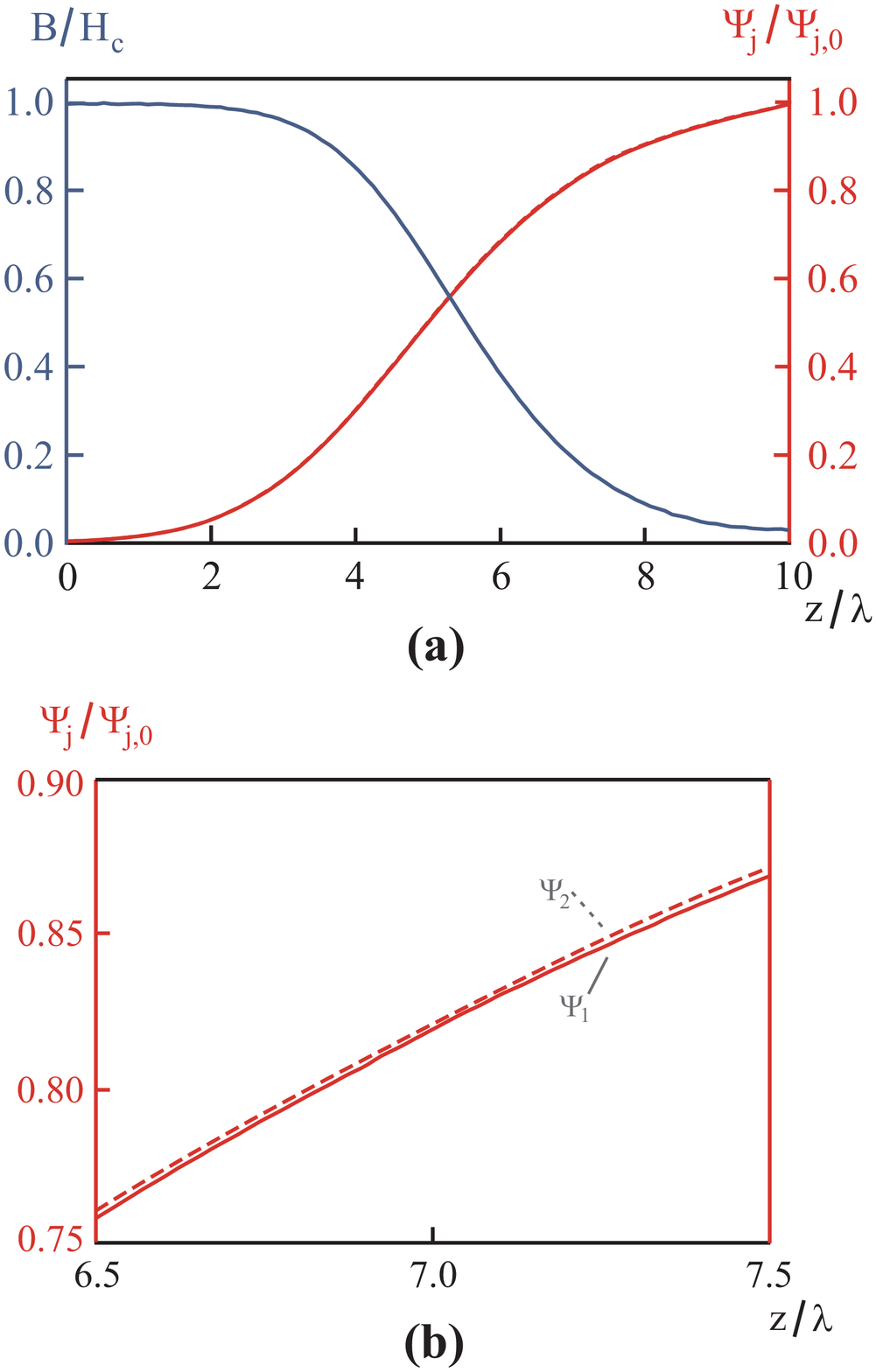}
\caption{(Color online) (a) The reduced field and order parameters obtained
numerically by minimizing the interface energy functional, $\Sigma$,
for $\kappa_{1}=0.45,\;\kappa_{2}=5,\; u=0.6,\; t=0.01$. The reduced
order parameter $\psi_{1}$ is shown by a solid line, the dashed line
is $\psi_{2}$. (b) The close-up of the order parameters for $6.5<z/\lambda<7.5$.}

\end{figure}

\begin{figure}
 \includegraphics[width=1\columnwidth]{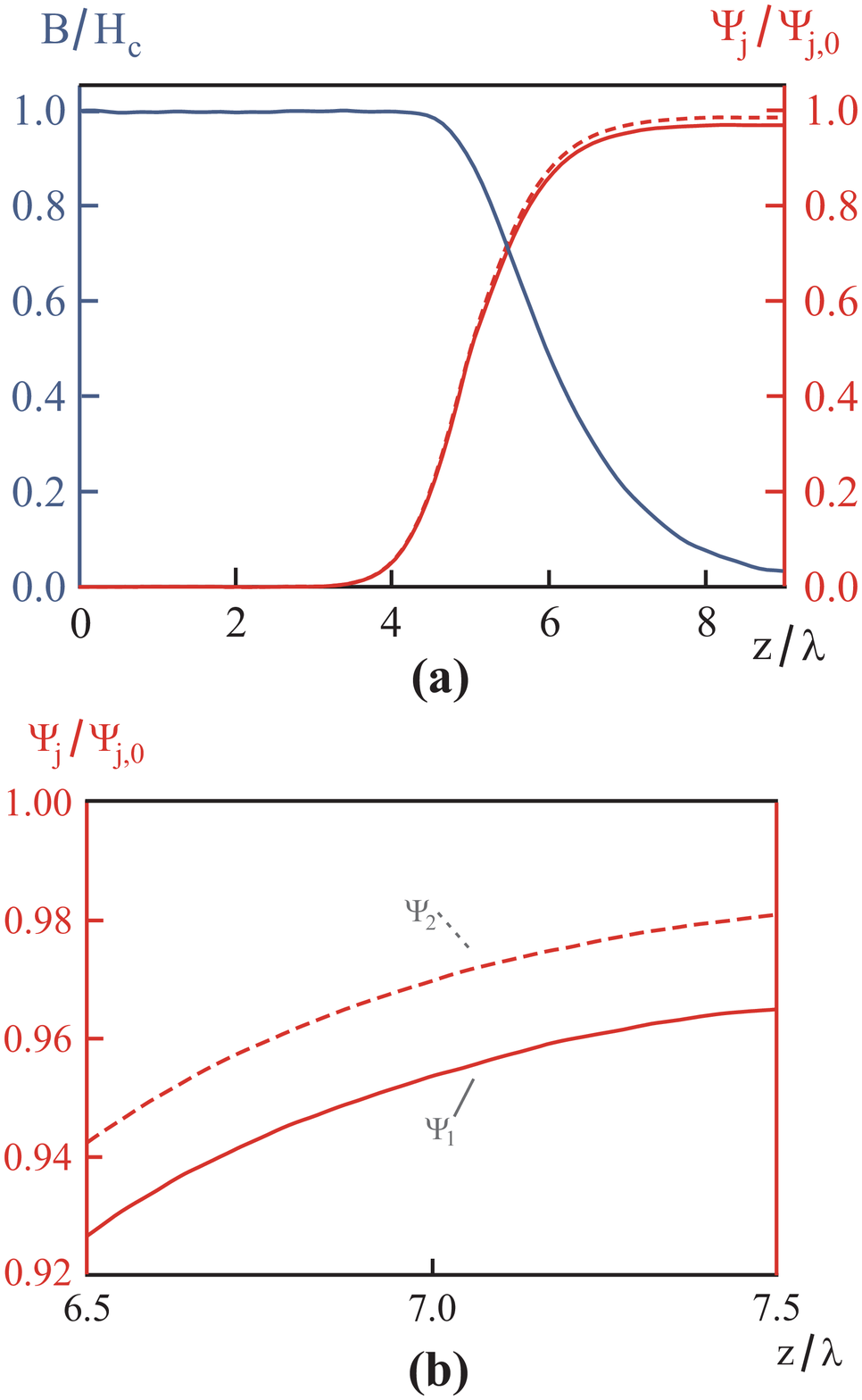}
\caption{(Color online) (a) The reduced field and order parameters obtained
numerically by minimizing the interface energy functional, $\Sigma$,
for $\kappa_{1}=3,\;\kappa_{2}=4,\; u=0.6,\; t=0.2$. The reduced
order parameter $\psi_{1}$ is shown by a solid line, the dashed line
is $\psi_{2}$. (b) The close-up of the order parameters for $6.5<z/\lambda<7.5$.}

\end{figure}

\begin{figure}
 \includegraphics[width=1\columnwidth]{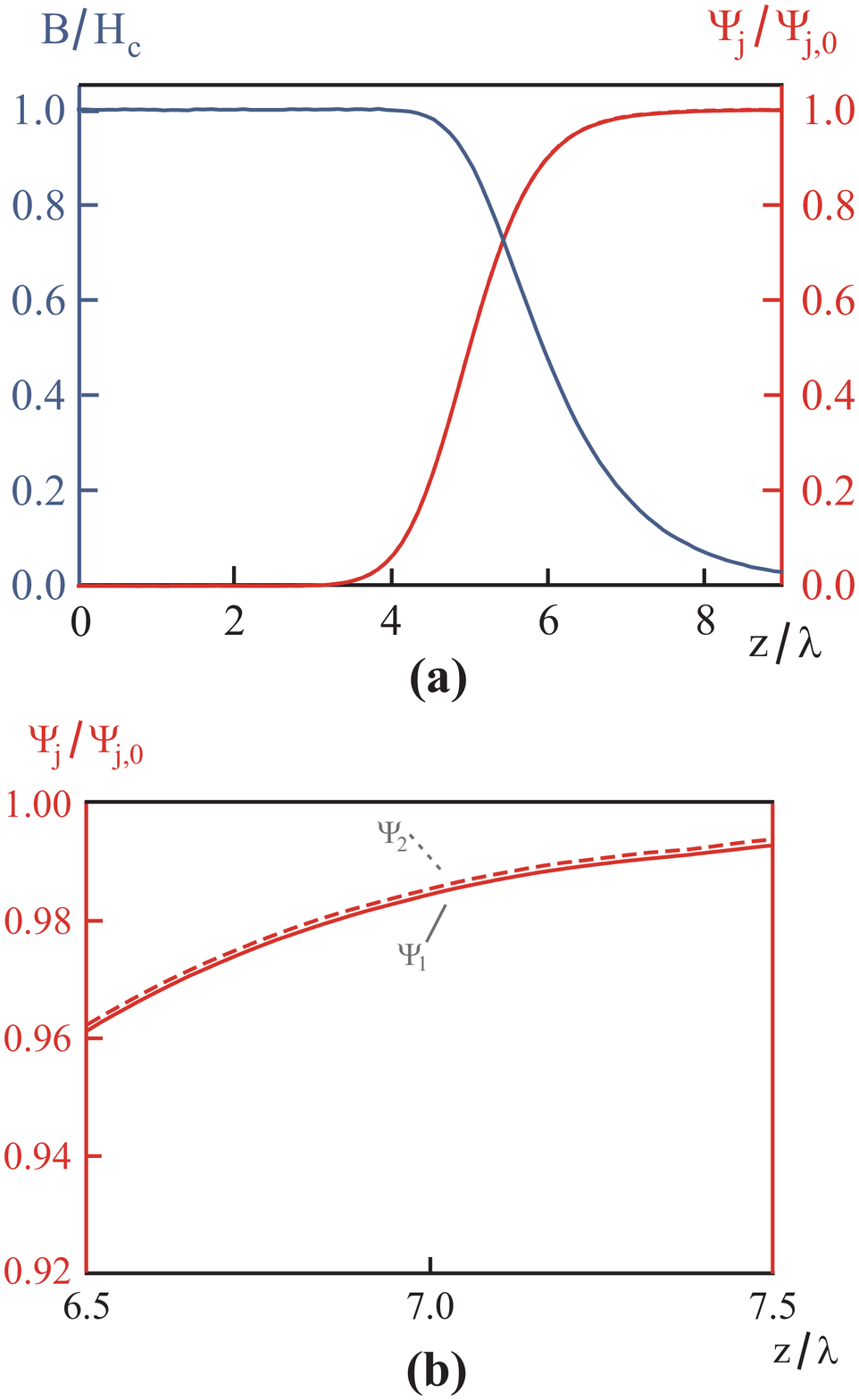}
\caption{(Color online) (a) The reduced field and order parameters obtained
numerically by minimizing the interface energy functional, $\Sigma$,
for $\kappa_{1}=3,\;\kappa_{2}=4,\; u=0.6,\; t=0.01$. The reduced
order parameter $\psi_{1}$ is shown by a solid line, the dashed line
is $\psi_{2}$. (b) The close-up of the order parameters for $6.5<z/\lambda<7.5$.}

\end{figure}

\section{Summary}

In summary, for a two-band superconductor we analyzed the energy
of the interface between regions of a finite order parameter and zero
order parameters, coexisting at the thermodynamic critical field $H_{c}$.
If one includes the interband Josephson coupling between the bands,
i.e. the leading allowed interaction between the Cooper-pair wave
function $\Psi_{1}$ and $\Psi_{2}$, both order parameters vary close
to the transition temperature on identical length scales. Thus, despite
the fact that both order parameters may have very different amplitudes,
they vary on the same characteristic length scale $\left(\xi_{1}^{-2}+\xi_{2}^{-2}\right)^{-1/2}$,
where the $\xi_{i}$ are the typical length scales where the gradient
(or kinetic) energies in the GL functional become comparable to the
bulk condensation energy. We stress that $\xi_{1,2}$ are just auxiliary
quantities and only $\xi$ is a measurable physical length. An important
implication of this result is that the surface energy is determined
by a single GL parameter $\kappa=\lambda/\xi$ in a way identical
to the single band case. 

Thus, there is no room left for so-called \emph{type-1.5} superconductivity
in the GL regime close to $T_{c}$. Of course, our analysis cannot
rule out the possibility of interesting novel physics due to distinct
charactetristic length scales $\xi_{i}$ deep in the superconducting
state. Such possibility then requires an approach within the microscopic
framework of the Bogoliubov-de Gennes or Gor'kov formalisms. In the
recent review by Brandt and Das,\cite{Brandt} situations are described
which do not fit to a rigid type-I- type-II dichotomy. Close to $T_{c}$,
however, only one relevant superconducting order parameter exists
and the phenomenology of the transition between type-I and type-II
superconductivity is unchanged by the multiband character of the system.

\section{Acknowledgements}

This work was supported by the U.S. Department of Energy, Office of
Basic Energy Sciences, DMSE. Ames Laboratory is operated for the U.S.
DOE by Iowa State University under Contract No. DE-AC02-07CH11358.
J. Geyer acknowledges support by the Harry Crossley Foundation as
well as NITheP.

\appendix

\section{Microscopic expression of the GL coefficients}

Within a weak coupling BCS theory it is possible to derive the parameters
of the GL expansion, Eqs.~(\ref{FGL})-(\ref{FGL3}), in terms of the microscopic
densities of states and pairing interactions:\cite{Zhitomirsky04}\begin{eqnarray}
a_{i} & = & N_{F}\left(\left(\lambda^{-1}\right)_{ii}-n_{i}\left(\ln\frac{2e^{\gamma}\omega_{D}}{\pi T_{c}}+\tau\right)\right),\notag\\
b_{i} & = & \frac{7\zeta\left(3\right)N_{F}}{8\pi^{2}T_{c}^{2}}n_{i},\notag\\
\eta & = & \frac{N_{F}}{\det\lambda}\lambda_{12}.\end{eqnarray}
 Here $N_{F}$ is the densities of states at the Fermi level per one
spin, $n_{i}=N_{F,i}/N_{F}$ are relative densities of states on two
bands, $\lambda_{ij}=N_{F}V_{ij}$ are interaction constants proportional
to the symmetric matrix $V_{ij}$ responsible for the Cooper pairing,\cite{comment1}
$\tau=\left(T_{c}-T\right)/T_{c}$, and $\left(\lambda^{-1}\right)_{11}=\lambda_{22}/\det\lambda$
etc. are elements of the matrix inverse to $\lambda_{ij}$.

The transition temperature follows from the condition $\left[a_{1}a_{2}\right]_{\tau=0}=\eta^{2}$
that leads to\begin{equation}
T_{c}=\frac{2e^{\gamma}}{\pi}\omega_{D}\exp\left(-1/\widetilde{\lambda}\right),\end{equation}
 with effective coupling constant\begin{eqnarray}
\widetilde{\lambda} & = & 2n_{1}n_{2}\mathrm{det}\lambda\left[n_{1}\lambda_{11}+n_{2}\lambda_{22}\right.\notag\\
 &  & -\left.\sqrt{\left(n_{1}\lambda_{11}-n_{2}\lambda_{22}\right)^{2}+4n_{1}n_{2}\lambda_{12}^{2}}\right]^{-1}.\end{eqnarray}
 It is now straightforward to express the variable $t$ in Eq.(\ref{t})
in terms of $\tau$ close to the transition temperature, which shows 
\begin{equation}
t=\frac{\det\lambda \sqrt{\left(n_{1}\lambda_{11}-n_{2}\lambda_{22}\right)^{2}+4n_{1}n_{2}\lambda_{12}^{2}}}{\lambda_{12}^{2}}\tau\end{equation}


\begin{thebibliography}{34}
\bibitem{comment1} Note, that the notation commonly used in the literature
$\lambda_{ij}^{\left(lit\right)}$ differs from $\lambda_{ij}$ used
here: $\lambda_{ij}^{\left(lit\right)}=n_{i}\lambda_{ij}$. In our
definition $\lambda_{ij}$ continues to be a symmetric (Hermitian)
matrix if $V_{ij}$ is symmetric (Hermitian).
\end{thebibliography}
\end{document}